\documentclass[12pt]{iopart}
\usepackage[pdftex]{graphicx,color}
\usepackage{amssymb}
\begin{document}
\title{Higher curvature counter terms cause the bounce in loop cosmology}
\author{Robert C. Helling}
\address{Arnold Sommerfeld Center\\ Ludwig-Maximilians-Universit\"at M\"unchen\\ Thereisenstra\ss e 37\\D-80333 M\"unchen\\Germany}
\ead{helling@atdotde.de}

\begin{abstract}
  In the loop approach to the quantisation of gravity, one uses a
  Hilbert space which is too singular for some operators to be
  realised as derivatives. This is usually addressed by instead using
  finite difference operators at the Planck scale, a process known as
  ``polymerisation''. In the symmetry reduced example of loop
  cosmology, we study an ambiguity in the regularisation which we
  relate to the ambiguity of fixing the coefficients of infinitely
  many higher curvature counter terms augmenting the Einstein-Hilbert
  action. Thus the situation is comparable to he one in a naive
  perturbative treatment of quantum gravity with a cut-off where the
  necessary presence of infinitely many higher derivative terms
  compromises predictability. As a by-product, we demonstrate in an
  appendix that it is possible to have higher curvature actions for
  gravity which still lead to first order equations of motion like in the
  Friedmann case.
\end{abstract}
\noindent Preprint LMU-ASC 58/09 

\def\mone{\mathbb{I}}

\section{Introduction}
Isotropic and homogeneous space-times not only model the universe at
the largest scales and are thus immediately relevant in cosmology but
provide an important truncation of any --- quantum or classical ---
theory of gravity. They are mini-superspace models in the sense that
the symmetry reduces the infinity of degrees of freedom of the
gravitational field to a finite number. As quantum theories, these are
quantum mechanical models which allow to ignore the additional
conceptual and technical complications of quantum field theories.  Thus
they provide ideal testing grounds to compare various approaches to
quantum gravity in a well controlled environment with as little of
notational clutter as possible. 

Classically, these space-times of Friedmann-Robertson-Walker type
start with a big bang singularity and a traditional canonical
quantisation \`a la Wheeler and deWitt does not change this picture. 

Recently, however, methods from loop quantum gravity have been applied
to this setting (see, for example \cite{Ashtekar:2006wn,
  Ashtekar:2006es, Ashtekar:2007em, Ashtekar:2007tv, Ashtekar:2006uz,
  Ashtekar:2006rx}). The surprising result was that the big
bang singularity is resolved into a big bounce. Looking backwards in time,
the universe only contracts to a maximum density and then
re-expands. In the case of a closed cosmology this leads to an infinite
oscillation of the universe between this bounce and the maximum
expansion when the gravitational pull stops the expansion and turns it
around.

In this letter, we will analyse the origin of the radically different
behaviour. We will find that it is due to a regularisation that is
characteristic of the loop approach: The kinematical Hilbert space is
based on an invariant state. This state, however, is so singular, it
does not allow canonical momenta to be represented by derivative
operators: The limit of infinitesimal translations does not exist. For
a discussion of alternatives of covariant states avoiding this
singular behaviour, see \cite{Helling:2004tb, Helling:2006yn,
  HellingTBA}.

The usual way to avoid this problem is to assume a minimal scale and
replace derivatives by finite difference operators of the order of the
Planck length $\ell_p$. In the full, non-reduced theory, this is
encountered as the fact that the spatial curvature is too singular to
be represented by a quantum operator but is re-expressed in terms of
holonomies around loops of Planck size which do have a quantum
representation. Here, in the cosmological setting, it means that
the canonical momentum is not a differential operator but the
difference between two translation operators. 

In \cite{Dzierzak:2009ip}, see also \cite{Malkiewicz:2009xz,
  Malkiewicz:2009qv}, it was shown that one can do the corresponding
  replacement already in the classical Hamilton function. The
  resulting ``polymerised'' equations of motion show the same bouncing
  behaviour as in the loop quantised cosmology. Therefore, strictly
  speaking, the bounce is not due to the quantisation but to the
  modification of the Hamiltonian (although that modification is
  motivated by quantum theory considerations).

There are, however, many different finite difference operators that
approximate the derivative in the limit $\ell_p\to 0$. Thus, there is no
canonical choice and indeed a lot of
arbitrariness\cite{Nicolai:2006id,Nicolai:2005mc}. All these different
finite difference operators differ by higher derivatives times
positive powers of $\ell_p$. In this note, we will investigate the
consequences of this arbitrariness. In conventional loop cosmology,
the momentum $p$ is replaced by $\sin(\ell_p p)/\ell_p$ but as we will
argue, any function $f(\ell_p p)/\ell_p$ with $f(0)=0$ and $f'(0)=1$
will do. We will show that as long as $f(p)$ has a local maximum at
some $p_{max}>0$ there will be a bounce exactly when $p=p_{max}$.

By Taylor expansion, different choices for $f$ differ by higher
derivative operators at 0. We will translate his ambiguity in a
covariant language and show that every choice of $f$ can be related to
a modification of the Einstein-Hilbert action by higher powers of the
curvature tensor. The loop cosmology choice of $f=\sin$ thus
corresponds to a particular, infinite collection of higher curvature
counter terms in the action. But as there is no physical principle
suggesting this particular choice of finite difference operator as
replacement for the momentum, any other choice of higher derivative
counter terms is as good as this.

We strongly believe that is conclusion not only holds in the symmetry
reduced cosmological theory but is characteristic for the loop
approach and thus also holds in the full four-dimensional field
theory: Different regularization of the spatial curvature in terms of
holonomies of loops of finite size again by higher derivatives upon
Taylor expansion. If the full theory is fully covariant, the terms in
this Taylor expansion are expressible in terms of covariant
derivatives of the curvature as those are the only tensors
available. Thus, one should once more expect that changes in the
regulator induce higher curvature terms in the action.

Our conclusion is therefore, that the situation regarding the
uniqueness (or rather non-uniqueness) of quantisation of gravity in
the loop framework is not much better than in a naive perturbative
field theory quantisation: There, finding Feynman loop integrals
diverging one can introduce a regularisation for example via a cut-off
(a role playing by the minimal scale $\ell_P$ in loop quantum gravity)
but then one has to face the problem of an effective theory that is to
determine an infinite choice of numerical coefficients for all
possible higher curvature counter terms. This freedom to choose a
counter term is parallelled in the choice of finite difference
operator or an expression for the spatial curvature in terms of
holonomies in the loop approach.

This note ends in an appendix which demonstrates that at every order
of the curvature tensor in the action, one can find a linear
combination of index contractions that leads to an action which
contains only first time derivatives of the scale factor and thus does
not lead to higher order equations of motion than in Einstein theory,
a result with some interest independent of the discussion of loop
cosmology.

Note added: After publication of the first version of this note, I was
made aware of \cite{DS}\ which has significant overlap with the work
presented here.

\section{Classical solution}
In the loop quantum gravity (LQG) literature, it is common to use an
ADM-type 
canonical 3+1 decomposition of the gravitational degrees of
freedom. Then, in the cosmological situation that we are interested in
as a toy mini-superspace model, one performs a symmetry
reduction. Here, however, for simplicity, we will start from a
Lagrangian formulation in terms of a metric of the FRW form with flat
spatial slices
\begin{equation}
  \label{eq:FRW}
  ds^2 = -dt^2 + a(t)^2 (dx^2+dy^2+dz^2).
\end{equation}
With the scale factor $a$, this contains a single degree of freedom
which only depends on the time coordinate $t$ that in our choice of
gauge measures proper time. This formalism leads to the identical
Hamiltonian treatment as the ADM plus symmetry reduction approach but
is more convenient when we later want to trace the influence of
higher curvature corrections to the Einstein-Hilbert action.

The only remains of coordinate invariance are shifts $t\mapsto t+c$
which are generated by the Hamiltonian (constraint) which will play a
central role later on. These shifts however make $a(t)$ not a good
observable since $t$ has no invariant meaning. To nevertheless have a
(local in time) observable degree of freedom one can use a relational
approach: One introduces a further scalar ``clock'' field $\phi$ to
the theory (which for simplicity we assume to be massless and
free). Then, in addition to $a(t)$ there is as well $\phi(t)$ which
turns out to be a monotonic function and thus can be inverted to
eliminate $t$. As a result, we obtain a true observable $a(\phi)$.

We will start from the Einstein-Hilbert action coupled to the scalar
suppressing all constants and using a natural system of units
\begin{equation}
  \label{eq:EHcov}
  {\cal S} = \int d^4x\sqrt{-g}(R -
  g^{\mu\nu}\partial_\mu\phi\partial_\nu\phi) = \int dt(-6 a\dot a^2 +
  a^3\dot\phi^2).
\end{equation}
It is important to realise that it would be wrong at this stage to
vary with respect to $a(t)$ and $\phi(t)$: This would lead to second
order equations of motion with a conserved energy. Doing so ignores
the Hamiltonian constraint implementing the time shifts. In fact, the
conserved energy is the Hamiltonian but rather finding it to be
conserved one should impose the condition that it vanishes. This
correct procedure leads to the Friedmann equation which is first order
in time while the second order equation above is the time derivative
of this equation.

This procedure is equivalent to varying the action before using the
ansatz to obtain the Einstein equations and only then to plug in the
metric ansatz. The Friedmann equation is then the 00-component of
Einstein's equation (the constraint) while the weaker second order
equations arise from the space-space components.

Before we do so, we will change variables: Since in loop quantum
gravity, it is the area which is naturally quantised in Planck units
we will instead of $a$ use $v=a^3$. In terms of this, the Lagrangian
reads
\begin{equation}
  \label{eq:Lagrangian}
  L= -\frac 23 \frac{\dot v^2}v + v\dot\phi^2.
\end{equation}
To proceed to a Hamiltonian formulation, we need the canonical momenta
\begin{equation}
  \label{eq:momenta}
  \beta = \frac{\partial L}{\partial \dot v} = -\frac 43 \frac{\dot
    v}v\qquad 
  p=\frac{\partial L}{\partial\dot\phi} = 2v\dot\phi  
\end{equation}
and obtain the Hamiltonian 
\begin{equation}
  \label{eq:clHamiltonian}
  H=-\frac 38 v\beta^2 +\frac 14 \frac{p^2}v = \frac 1v
  \left(\sqrt{\frac 38}v\beta +\frac 12 p\right)
  \left(-\sqrt{\frac 38}v\beta +\frac 12 p\right).
\end{equation}
This, as explained above, leads to the constraint
$\sqrt{3/8}v\beta\approx \pm p/2$ (as usual equations with $\approx$
hold on the constraint surface). We have written the Hamiltonian in
this factorised form as $v$ is always positive and thus depending on
the signs of $\beta$ and $p$ only one of the factors can
vanish. Swapping the two factors corresponds then to time
reversal. This factorisation is useful since if $H=H_1H_2$ and in fact
$H_1\approx 0$, one can express the time evolution of an phase space
function $A$ as $\{H,A\}\approx \{H_1,A\}H_2$. As long as one is only
interested in ratios of time derivatives of phase space functions one
can thus use $H_1$ instead of $H$ as we will do repeatedly below. 

The Hamilton equations of motion
\begin{equation}
  \label{eq:eom}
  \dot p=0\qquad \dot\phi=\frac p{2v}\qquad \dot v = -\frac
  3v\beta\approx \pm\sqrt{3/8}p
\end{equation}
are solved by $v=|\sqrt{3/8} p(t-t_0)|$ and $\phi=  \sqrt{2/3}\ln|t-t_0|$. This
reproduces the expected big bang/big crunch singularity $v=0$ at finite
proper time $t=t_0$.

The evolution of the volume as a function of the clock field $v=
\exp(\pm \sqrt{3/2}(\phi-\phi_0))$ has the singularity at $\phi=\pm\infty$. It is
either found algebraically from the solutions $v(t)$ and $\phi(t)$ or
directly by solving the gauge invariant equation of motion
\begin{equation}
  \label{eq:gaugeeom}
  \frac{dv}{d\phi} = \frac{\dot v}{\dot \phi} = \frac{\partial
    H/\partial\beta} {\partial H/\partial p}= -\frac{3\beta
    v^2}{2p}\approx\pm\sqrt{3/4}v. 
\end{equation}
This concludes our discussion of the classical system
(\ref{eq:EHcov}).

\section{The LQG modification}
The Hamiltonian (\ref{eq:clHamiltonian}) is simple enough to be
quantised canonically by considering wave functions $\Psi(v,\phi)$ and
replacing $v$ and $\phi$ by multiplication operators and the momenta
by derivatives $\hat \beta = -i\partial/\partial v$ and $\hat p =
-i\partial/\partial\phi$. This Wheeler-deWitt equation can then be
turned into a wave equation by the substitution $x=\ln(v)$.

This simple quantisation is, however, not available in the context of
loop quantum gravity. There, the kinematical Hilbert space is
constructed as the GNS representation built upon an invariant state
(for an extended discussion, see for example \cite{Helling:2004tb,
  Helling:2006yn, HellingTBA}. The invariance (as opposed to
covariance as in the Wheeler-deWitt theory) implies a
discontinuous representation of at least one of the unitary operators
which can be thought of the exponentials of the canonical variables
$v$ and $\beta$. 

As a result, the derivative $-i \partial/\partial v$ is too singular
to be defined in this ``polymer Hilbert space''. In the full,
covariant theory this is encountered as the problem that the spatial
curvature is too singular to be defined as an operator (as it would be
a derivative in a conventional treatment). 

The solution found in the loop literature is to observe that although
the curvature does not exist as a quantum operator, the holonomies do
and the curvature can be approximated by holonomies around small
loops. In the reduced theory discussed here, this translates to the
observation that although the momentum $\beta$ cannot be defined as a
quantum operator in the polymer space, the translation operator for a
finite distance in $v$ can via $U(\ell)\Psi(v,\phi) =
\Psi(v+\ell,\phi)$. This operator can be viewed as the quantisation of
the Weyl operator $\exp(i\ell \beta)$ which does not suffer the
problems of the quantisation of $\beta$.

In the full theory, it is then appealed to foaminess of space-time in
a theory of quantum gravity and any occurrence of the spatial
curvature is replaced by an expression in terms of holonomies around
loops of Planck size which classically converge to the curvature in
the limit of zero loop size. As in the quantum theory the limit does
not exist one is satisfied with finite size loops of Planck length.

To parallel this procedure, in loop quantum cosmology, the occurrence
of the momentum $\beta$ in the Hamiltonian is replaced by the operator
\begin{equation}
  \label{eq:betatilde}
  \tilde\beta=\frac{U(\ell)-U(-\ell)}{2i\ell}.  
\end{equation}
This replacement is know as ``polymerisation'', see for example
\cite{Kunstatter:2008qx}.  Note well, that this is the quantisation of
the classical $\sin(\ell\beta)/\ell$.
 
The distance $\ell$ is of
course kept fixed as the limit $\ell\to 0$ does not exist in the
quantum theory. This is motivated by observing that in the classical
theory,
$\lim_{\ell\to0}(\exp(i\ell\beta)-\exp(-i\ell\beta))/2i\ell = \beta$.
As in the numerical solution of differential equations, the derivative
$\hat\beta=\-i\partial/\partial v$ is replaced by the finite
difference operator $\tilde\beta$.

At this point, the change of variables from $a$ to $v=a^3$ is
important: The operator $\tilde\beta$ relates the wave function at two
different points of constant difference $\Delta v=\ell$ rather than at
constant $\Delta a$. We see that the regularisation explicitly breaks
coordinate independence and singles out the coordinate $v$.

The regularisation is, however, far from unique. In a section below,
we will study the consequences of this arbitrariness. 

In this LQC modified quantum theory, it is then found
\cite{Ashtekar:2006wn, Ashtekar:2006es, Ashtekar:2007em,
  Ashtekar:2007tv, Ashtekar:2006uz, Ashtekar:2006rx} that the time
evolution is radically different compared to the classical theory
investigated above: The contraction (looking back in time) is is
stopped at a finite value for $v$ and $\beta$ and, instead of a big
bang or big crunch, there is a bounce and the universe re-expands. 

As explained in \cite{Dzierzak:2009ip}, see also
\cite{Malkiewicz:2009xz, Malkiewicz:2009qv}, one does not have to
(possibly numerically) solve the quantum time evolution equation (for
effective actions used in loop cosmology, see also
\cite{Bojowald:2006ww,Bojowald:2007cd,Sotiriou:2008ya}). It is
sufficient to replace $\beta$ by $\sin(\ell\beta)/\ell$ in the
classical Hamiltonian, viz
\begin{equation}
  \label{eq:LQCHamiltonian}
  \tilde H = -\frac 38 \frac{v\sin(\ell\beta)^2}{\ell^2} +
  \frac{p^2}{4v}= \frac 1v
  \left( \sqrt{\frac 38}\frac{v\sin(\ell\beta)}\ell +\frac p2\right)
  \left(-\sqrt{\frac 38}\frac{v\sin(\ell\beta)}\ell +\frac p2\right)
\end{equation}
and solve the resulting equations of motion. In particular, for the
gauge invariant evolution, one now finds
\begin{equation}
  \label{eq:LQCgaugeeom}
  \frac{dv}{d\phi} = \frac{\dot v}{\dot \phi} = \frac{\partial
    H/\partial\beta} {\partial H/\partial p}= \sqrt{\frac
    32}v\cos(\ell\beta)\approx \pm \sqrt{\frac 32 v^2-(\ell p)^2}.
\end{equation}
This is solved by $v=\sqrt{3/2} \ell |p|
\cosh(\sqrt{3/2}(\phi-\phi_0))$. Obviously, this solution is a
bouncing universe with a minimum at $v_{min}= \sqrt{3/2}\ell|p|$. Note
well that the momentum (curvature) at the minimum is $\beta_{max} =
\pi/2\ell$ independent of initial conditions. Going back to the
equation of motion, we see that the universe bounces when $\dot
v=\partial H/\partial\beta=0$.
\begin{figure}[h]
  \centering
  \includegraphics[width=0.6\textwidth]{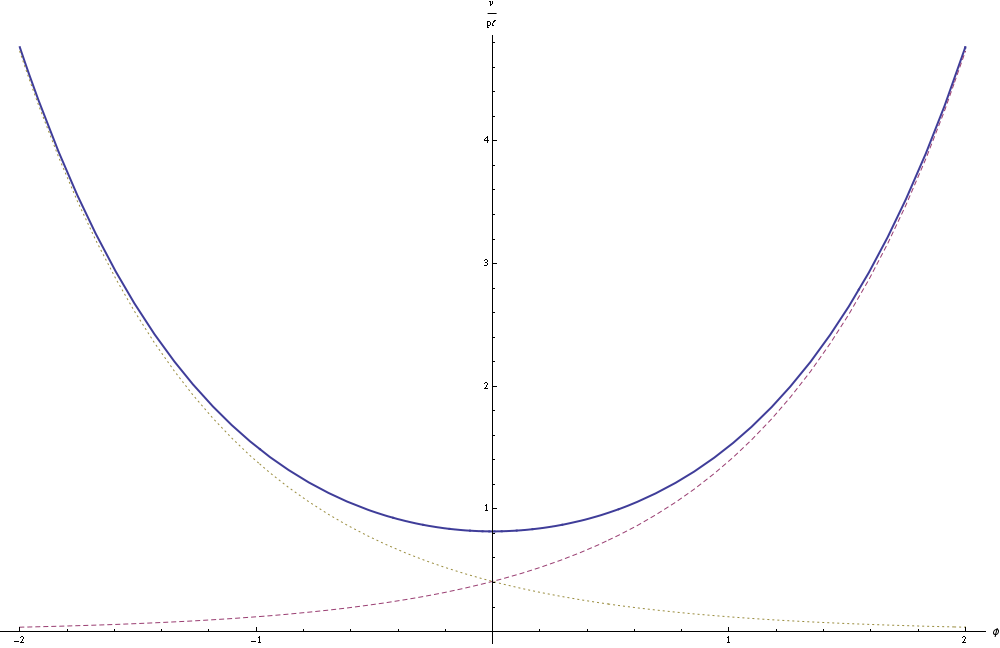}
  \caption{Evolution of the scale factor: LQC bounce (solid),
    classical expansion (dashed) and classical contraction (dotted)}
  \label{fig:bounce}
\end{figure}

We conclude that the bouncing behaviour characteristic for loop quantum
cosmology can already be found in the classical evolution once one
applies the LQG substitution $\beta\mapsto \sin(\ell \beta)/\ell$ in
the Hamilton function. One can then obtain the LQC wave equation (the
finite difference equation) form the Wheeler-de Witt-quantisation of
the LQG-modified Hamiltonian as $\exp(\ell\partial_v)$ is in fact the
translation operator $\Psi(v)\mapsto\Psi(v+\ell)$. Thus we can view
loop quantum cosmology as an ordinary quantum theory with a modified
Hamiltonian.

\section{Counter term ambiguities}
As it is well known from numerical evaluation of derivatives, the
replacement of the derivative with respect to $v$ by the finite
difference operator (\ref{eq:betatilde}) is far form unique. This was
already stressed in \cite{Nicolai:2006id,Nicolai:2005mc}, lattice
refiniement for loop cosmology was also addressed in
\cite{Bojowald:2007qu}. 
One could
as well use 
\begin{equation}
  \label{eq:beta2tilde}
  \tilde\beta'=\frac{U(2\ell)-U(-2\ell)}{4i\ell}  
\end{equation}
or for example a linear combination
\begin{equation}
  \label{eq:beta12tilde}
  \tilde\beta_\alpha=\alpha\frac{U(\ell)-U(-\ell)}{2i\ell}+(1-\alpha)\frac{U(2\ell)-U(-2\ell)}{4i\ell}  
\end{equation}
or a linear combination of even more translation operators. All these
have in common that the Taylor expansion for small $\beta$ starts with
$\beta+O(\ell^2\beta^3)$. For example, we have
$\tilde\beta_\alpha=\beta+\frac{3\alpha-4}6 \ell^2\beta^3 +
O(\ell^4\beta^5)$. In general, with a the appropriate choice of $n$
translation operators $U$, one pick the first $n$ terms in the Taylor
expansion for small $\beta$. 

Therefore, with arbitrary precision for small $\beta$ (for example up
to a possible maximum at a bounce, see below), one can approximate any
function $f(\ell\beta)/\ell$. The only physical condition is the
$\beta\to 0$ limit (the ``unloopy limit'') in which it should be given
by $\beta$, that is $f(0)=0$ and $f'(0)=1$. 

More generally, one could also use an expression that also depends on
the other variables $v$, $\phi$, and $p$. An example would be the
above mentioned possibility to consider constant step sizes in $a$
rather than $v$. In this letter, we will for simplicity, however,
restrict our attention to functions $f$ which only depend on $\beta$.

For general such $f$, the Hamiltonian reads
\begin{equation}
  \label{eq:fHamiltonian}
  H_f = -\frac 38 \frac{vf(\ell\beta)^2}{\ell^2} +
  \frac{p^2}{4v}= \frac 1v
  \left( \sqrt{\frac 38}\frac{vf(\ell\beta)}\ell +\frac p2\right)
  \left(-\sqrt{\frac 38}\frac{vf(\ell\beta)}\ell +\frac p2\right)  
\end{equation}
the gauge invariant evolution equation
corresponding to (\ref{eq:gaugeeom}) and (\ref{eq:LQCgaugeeom}) now
reads 
\begin{equation}
  \label{eq:generalgaugeeom}
    \frac{dv}{d\phi} = \frac{\dot v}{\dot \phi} = \frac{\partial
    H/\partial\beta} {\partial H/\partial p}= 
  \sqrt{\frac
    32}vf'(\ell\beta).
\end{equation}
We see that there is a bounce when $f'(\ell\beta)=0$ for some
$\beta$ whereas there is a big bang/big crunch when $f$ is strictly
monotonic. For example, the classical case (\ref{eq:clHamiltonian})
corresponds to $f(x) = x$ and has an initial singularity whereas the
LQC case with $f=\sin$ bounces.

Coming from a loop perspective, one would restrict attention to
functions $f$ that can be build out of Weyl operators $U(\cdot)$ and
symmetry then restricts one to linear combinations $f(x) = \sum_n a_n
\sin(b_n x)$. Obviously, finite sums always have maxima and thus
always bounce. But even in the polymer Hilbert space, one can have
infinite sums as long as $\sum_n|a_n|^2<\infty$. Amongst those there
is for example 
\begin{equation}
  \label{eq:singular}
  f_s(x) =  \frac1{(32^{1/4}-1)\zeta(-1/4)}\sum_{n=1}^\infty\frac{(-1)^n}{n^{3/4}}\sin(nx),
\end{equation}
which is monotonic between $-\pi/2<x<\pi/2$ and has poles at
$|x|=\pi/2$. The time evolution with this polymerisation function is
even more singular than in the classical case as it reaches $v=0$ for
finite $\phi$.

All possible choices for $f$ will differ by higher powers of $\ell\beta$ which
in the conventional quantisation corresponds to higher derivatives
with respect to $v$ with coefficients that vanish for $\ell=0$. In the
remainder of this section, we will argue that these higher derivative
terms can be translated back to the action where they correspond to
the addition of higher curvature corrections to the Einstein-Hilbert
action. Thus, any choice for a regularising function $f$ corresponds
to a particular choice of higher curvature counter terms in the
action.

One should compare this to what one would do in a naive perturbative
quantisation of gravity: There, one would expand the Einstein-Hilbert
action around some background to obtain Feynman rules and then would
find that the Feynman integrals are divergent and the theory is
non-renormalisable since Newton's constant which controls the
perturbative expansion has negative mass dimension. All integrals,
however, are finite if one uses some regularisation (for example
point-splitting where operators cannot get closer than a distance
$\ell$). But then, one should treat gravity as an effective theory and
not only include the Einstein-Hilbert term but all possible scalars
that can be built out of powers of covariant derivatives of the
curvature. An infinite number of coefficients would have to be
determined experimentally and would be subject to change once one
changes the regulator. 

We see that here the situation is very similar: By picking a
regularising function $f$ which avoids short distance singularities as
in a differential quotient representing $\beta$ in traditional quantum
mechanics, one implicitly chooses a particular set of coefficients for
the higher curvature counter terms. The ambiguity in $f$ can be mapped
to the well known ambiguity in picking a higher derivative effective
action.

To see this, let us first translate the LQG corrected Hamiltonian
(\ref{eq:fHamiltonian}) back to a Lagrangian $L= \dot v\beta +
\dot\phi p -H_f$. Unfortunately,
 this cannot be done explicitly for 
general $f$ since one has to invert $\dot v= -(3/4) v f(\ell \beta)
f'(\ell\beta)/\ell$ to find $\beta$ as a 
function of $\dot v$. Thus, 
we present it here in the LQC case $f=\sin$. We find $\beta =
-\arcsin(8 \ell \dot v/3 v)/2 \ell$ 
and eventually  
\begin{equation}
  \label{eq:Lbar}
  \bar L= \frac{3a^3}{16\ell^2} \left( 1-\sqrt{1-s^2} - s\arcsin
  (s)\right)+a^3\dot\phi^2, 
\end{equation}
where we defined $s=8\ell(\dot a/a)$ which is related to the square
root of the curvature. This Lagrangian contains arbitrary even powers
of the derivative $\dot a$ and, in the limit $\ell\to0$, reduces to
(\ref{eq:Lagrangian}). It was already discussed in \cite{DS}.

As a final step, we want to discuss how such a higher derivative
Lagrangian can arise from a higher curvature covariant action in four
dimensions. There is, 
however, a minor technical difficulty: For the
metric (\ref{eq:FRW}), the Ricci scalar is
\begin{equation}
  \label{eq:Ricci}
  R=6(\dot a /a)^2-6 \ddot a/a,
\end{equation}
and the Einstein-Hilbert Lagrangian ${\cal L}=\sqrt{-g}R$ not only
contains $\dot a$ but as well the second derivative $\ddot a$. So far,
this did not cause any trouble when proceeding to the Hamiltonian
setting, since one can integrate by parts to get rid of second time
derivatives. If one naively adds higher powers of the Ricci scalar
$R^n$ to the action, not all second derivatives $\ddot a$ can be
eliminated by integration by parts anymore and thus (\ref{eq:Lbar})
can not be written covariantly only in terms of the Ricci scalar. It
is well known in the literature, that the generalisation of the
Friedmann equation including generic higher order curvature correction
ceases to be first order.

This difficulty can be avoided once one includes traces of higher
powers of the Ricci tensor $R^{(n)} =
R_{\mu_1}{}^{\mu_2}R_{\mu_2}{}^{\mu_3}\cdots R_{\mu_n}{}^{\mu_1}$.
In the appendix, it is shown that up to a total derivative, any
function $a^3P((\dot a/a)^2)$ for analytic $P$ can be written as
$\sqrt{-g}$ times a power series in the $R^{(n)}$. In particular, this
is true for the LQC Lagrangian $\bar L$ above (cf.~(\ref{eq:Lbar})).

We1 have therefore shown that the polymerisation of the Hamilton
operator by replacing $\beta$ by $f(\ell \beta)/\ell$ is equivalent to
a modification of the Einstein Hilbert action by higher order
curvature corrections. Different choices for the function $f$
correspond to different choices of higher derivative counter terms.

This can be compared to a naive field theoretic quantisation of
gravity: Even though the Einstein theory is non-renormalisable, one
can still expand around a fixed background and compute Feynman loop
integrals. Those, of course, will be divergent but can be regularised
for example by point splitting. To balance the effect of changing the
regulator one then has to add an infinity of counter terms with
coefficients to be determined by experiment. Any choice of counter
terms is as good as any other as an effective theory. 

We find exactly the same situation in loop quantum cosmology: One
evades short distance singularities by point splitting in time
(replacing derivatives by finite difference operators with step size
of order $\ell$) and this induces an infinite series of higher
curvature corrections. Once more, the regularisation is far from
unique and changing it changes the numerical coefficients of the
counter terms.

Thus the finiteness properties of loop cosmology are just the same as
in naive field theoretic quantisation with a finite cut-off (or point
splitting). Presumably this conclusion extends to the full, symmetry
unreduced theory of loop quantum gravity.

\section{Conclusions}
Perturbative quantum gravity in four dimensions is known to be
non-renormalisable. Still, it can be made finite by hand by regulating
divergent loop integrals, for example by point splitting. Slight
changes in the regulator will induce changes of the coefficients of an
infinity of higher curvature terms in the effective action that have
positive powers of the cut-off scale as coefficients. This suggests
that those higher curvature counter terms should have been included
in already at the classical action and their coefficients eventually
have to be determined experimentally as they are not predicted by the
theory. Such a treatment of quantum gravity is not satisfactory due to
its lack of predictability.

Loop quantum gravity is a non-perturbative approach to the
quantisation of gravity supposedly avoiding these problems from
non-renormalisability by manifest finiteness as at no stage divergent
Feynman integrals have to be computed. In this note, we have given
evidence that this not the case and in fact the situation in loop
quantum gravity is the same as in the naive, perturbative treatment.
We have shown how he same problems of ambiguities in higher curvature
terms reappear also in the loop approach. 

There, the ``polymer'' Hilbert space does not allow to define
operators for the spatial curvature as the derivative with respect to
the holonomies that it would be in conventional quantisation is too
singular to exist. Therefore, the curvature has to be approximated by
holonomies around Planck sized loops. Classically, this approximation
would be exact in the limit of shrinking the loop to a point, but in
the quantum theory it is kept constant at the Planck scale. This step was
argued to be the analog of regularisation and is well known  not to be
unique. 

In this note, we studied the ambiguity of this regularisation:
There is an infinity of inequivalent expression in terms of holonomies
that approximate the curvature once the loops shrink to points. 

We studied this in detail in the symmetry reduced version of loop
quantum gravity known as loop cosmology. There, the scale factor $a$
and a scalar field $\phi$ only depend on time and instead of
representing the canonical momentum for $a$ (or the volume element
$v=a^3$) by a derivative one regulates it by a finite difference
operator at lattice spacing proportional to $\ell$. 

This ``polymerisation'' can already be performed at the classical
level by replacing the momentum $\beta$ by
$\sin(\ell\beta)/\ell$. Once more, this regularisation is far from
unique and the ambiguity can be parametrised by a function $f$ with
$f(0)=0$ and $f'(0)=1$. Any such $f$ leads to a different
regularisation. 

We then translated this ambiguity back to the level of the action
where we demonstrated that different $f$ correspond to the addition of
different higher curvature corrections to the Einstein-Hilbert
action. Therefore, in the loop treatment, it seems, there is the same
lack of predictability due to an infinite number of possible
counter-terms with unknown coefficients as in the naive, perturbative
approach. This has not been considered so far since only a single
polymerisation function $f=\sin$ has been considered in the
literature. Unfortunately, there is no physical principle that singles
out this particular choice and others should be regarded as equally
well motivated as this.

We used the fact that the regularisation can already be applied to the
classical theory before quantization although the modification is
only required after quantization to give well defined operators in
the appropriate (polymer) Hilbert space. This modified classical
theory, however, already shows features that are supposedly quantum in
nature like a bounce avoiding  a big bang/big crunch singularity. By
restricting our attention to this modified classical theory we could
avoid the question if the full quantised theory actually exists and if
it can be described by a simple effective action as we have been using
it. 

\ack
The author would like to thank John Baez, Giuseppe Policastro, and
Ivo Sachs for discussions and suggestions and Ghanashyam Date for pointing out \cite{DS}. This work was supported by
the Elitenetzwerk Bayern. 

\section*{Appendix}
In this appendix, we will show that it is possible to write every even power
of $(\dot a/a)^{2n}$ as a linear combination of contractions of $n$
powers of the Ricci tensor up to a total time derivative. Thus, every
effective Lagrangian that arises by polymerising with some
function $f$ can be obtained from a covariant four-dimensional
Lagrangian. 

This result is of independent interest since it shows that it is
possible to have gravity with higher curvature corrections while
maintaining the first order nature of the corrected Friedmann equation.

The problem arises as the Ricci scalar (\ref{eq:Ricci}) alone not only
contains $(\dot a/a)^2$ but a second derivatives $\ddot a/a$ as well. A
term that contains only one factor $\ddot a$ can be converted to a
term containing only $a$ and $\dot a$, but this is not possible for
higher powers $\ddot a^k$. We will show, that at each power $n$ of the
curvature, it is possible to find a linear combination of contractions
of the Ricci tensor (as the metric (\ref{eq:FRW}) is conformally flat,
the Weyl tensor vanishes) that contains only $(\dot a/a)^{2n}$ as well
as $(\ddot a/a) (\dot a/a)^{2n-2}$ which can be converted to the
wanted term by integration by parts.

It is convenient to go the conformal frame by changing to a new time
coordinate $T$ defined by $dT/dt=1/a$ in which the metric reads
\begin{equation}
  \label{eq:confmetric}
  ds^2=a^2(-dT^2+dx^2+dy^2+dz^2).
\end{equation}
In this metric, the Ricci tensor is 
\begin{equation}
  \label{eq:FRWRicci}
  R_\mu{}^\nu = \pmatrix{3 (a'/a^2)^2-3a''/a^3&\cr
    &-\left((a'/a^2)^2+ a''/a^3\right)\mone_3\cr
  }
\end{equation}
where we denote $a'=da/dT=a\dot a$ and $a''=d^2a/dT^2=a\dot
a^2+a^2\ddot a$. As before, we denote traces of powers of the Ricci
tensor by $R^{(n)} = R_{\mu_1}{}^{\mu_2}R_{\mu_2}{}^{\mu_3}\cdots
R_{\mu_n}{}^{\mu_1}$. $R^{(1)}= -6a''/a^3$ is the Ricci scalar and
does not contain the ``wanted term'' $(a'/a^2)^2=(\dot a/a)^2$ without
second derivatives.

Let us write $A_{k,l}$ for $(a'/a^2)^{2k}(a''/a^3)^l$ with
$k,l\ge0$. We want to show that for each $n$ there is a polynomial
$P_n$ in the $R^{(k)}$ which is a linear combination of only $A_{n,0}$
and $A_{n-1,1}$. For $n>1$, $P_n$ is not a multiple of $A_{n-1,1}$.

We prove this by induction on $n$. For $n=1$, this is true since
$R^{(1)}= -6A_{0,1}$. Now assume, we found the $P_l$ for all $l<n$.
We start with $R^{(n)}$, this contains $A_{n,0}$ with coefficient
$3^n+3(-1)^n$ which does not vanish for $n>1$. In general, it is a
linear combination of $n+1$ terms $A_{n-k,k}$ with $0\le k\le n$. Now
observe that, by induction, $P_k^{(n)}:=P_k {R^{(1)}}^{n-k}$ contains
only $A_{k,n-k}$ and $A_{k-1,n-k+1}$ for $n-1$ values
$k=1,\ldots,n-1$. These form a basis of the vector space spanned by
$\{A_{n-2,2},A_{n-3,3},\ldots,A_{0,n}\}$. Thus it is possible to find
$P_n$ as a linear combination of $R^{(n)}$ and the $P_k^n$.

This completes our proof, since $A_{n,0}= (\dot a/a)^{2n}$ and
$A_{n-1,1}=(\dot a/a)^{2n}+(\dot a/a)^{2n-2}\ddot a/a$ and since the second
term can be converted to $(\dot a/a)^{(2n)}$ as well by an integration
by parts up to a total derivative.

\section*{References}
\bibliography{lqc}
\bibliographystyle{utphys}

\end{document}